\documentclass[twoside,twocolumn,9pt]{article}
\usepackage{extsizes}
\usepackage[super,sort&compress,comma]{natbib} 
\usepackage[version=3]{mhchem}
\usepackage[left=1.5cm, right=1.5cm, top=1.785cm, bottom=2.0cm]{geometry}
\usepackage{balance}
\usepackage{times,mathptmx}
\usepackage{sectsty}
\usepackage{graphicx} 
\usepackage{lastpage}
\usepackage[format=plain,justification=justified,singlelinecheck=false,font={stretch=1.125,small,sf},labelfont=bf,labelsep=space]{caption}
\usepackage{float}
\usepackage{fnpos}
\usepackage[english]{babel}
\usepackage{array}
\usepackage{droidsans}
\usepackage{charter}
\usepackage[T1]{fontenc}
\usepackage[usenames,dvipsnames]{xcolor}
\usepackage{setspace}
\usepackage[compact]{titlesec}
\usepackage{hyperref}

\usepackage{epstopdf}

\definecolor{cream}{RGB}{222,217,201}
\newcommand{\SI}{Supplemental~Material}
\begin{document}



\makeFNbottom
\makeatletter
\renewcommand\LARGE{\@setfontsize\LARGE{15pt}{17}}
\renewcommand\Large{\@setfontsize\Large{12pt}{14}}
\renewcommand\large{\@setfontsize\large{10pt}{12}}
\renewcommand\footnotesize{\@setfontsize\footnotesize{7pt}{10}}
\renewcommand\scriptsize{\@setfontsize\scriptsize{7pt}{7}}
\makeatother

\renewcommand{\thefootnote}{\fnsymbol{footnote}}
\renewcommand\footnoterule{\vspace*{1pt}%
\color{cream}\hrule width 3.5in height 0.4pt \color{black} \vspace*{5pt}} 
\setcounter{secnumdepth}{5}

\makeatletter 
\renewcommand\@biblabel[1]{#1}            
\renewcommand\@makefntext[1]%
{\noindent\makebox[0pt][r]{\@thefnmark\,}#1}
\makeatother 
\renewcommand{\figurename}{\small{Fig.}~}
\sectionfont{\sffamily\Large}
\subsectionfont{\normalsize}
\subsubsectionfont{\bf}
\setstretch{1.125} 
\setlength{\skip\footins}{0.8cm}
\setlength{\footnotesep}{0.25cm}
\setlength{\jot}{10pt}
\titlespacing*{\section}{0pt}{4pt}{4pt}
\titlespacing*{\subsection}{0pt}{15pt}{1pt}


\makeatletter 
\newlength{\figrulesep} 
\setlength{\figrulesep}{0.5\textfloatsep} 

\newcommand{\topfigrule}{\vspace*{-1pt}%
\noindent{\color{cream}\rule[-\figrulesep]{\columnwidth}{1.5pt}} }

\newcommand{\botfigrule}{\vspace*{-2pt}%
\noindent{\color{cream}\rule[\figrulesep]{\columnwidth}{1.5pt}} }

\newcommand{\dblfigrule}{\vspace*{-1pt}%
\noindent{\color{cream}\rule[-\figrulesep]{\textwidth}{1.5pt}} }

\makeatother

\twocolumn[
  \begin{@twocolumnfalse}
\vspace{3cm}
\sffamily

 \noindent\LARGE{\textbf{Spin-State dependent Conductance Switching in Single Molecule-Graphene Junctions$^\dag$}} \\
 \vspace{0.2cm} \\

 \noindent\large{Enrique Burzur\'i,$^{\ast}$\textit{$^{a,b}$} Amador Garc\'ia-Fuente,\textit{$^{c}$} Victor Garc\'ia-Su\'arez,\textit{$^{c}$} Kuppusamy Senthil Kumar,\textit{$^{d,e}$} Mario Ruben,\textit{$^{d,e}$} Jaime Ferrer,$^{\ast}$\textit{$^{c}$} and Herre S. J. van der Zant \textit{$^{a}$}} \\

 \end{@twocolumnfalse} \vspace{1.5cm}

  ]

\renewcommand*\rmdefault{bch}\normalfont\upshape
\rmfamily
\section*{}
\vspace{-1cm}


\footnotetext{\textit{$^{a}$~Kavli Institute of Nanoscience, Delft University of Technology, PO Box 5046, 2600 GA Delft, The Netherlands. }}
\footnotetext{\textit{$^{b}$~IMDEA Nanoscience, Ciudad Universitaria de Cantoblanco, C/Faraday 9, 28049 Madrid, Spain; E-mail: enrique.burzuri@imdea.org}}
\footnotetext{\textit{$^{c}$~Departamento de F\'isica, Universidad de Oviedo and CINN (CSIC), ES-33007 Oviedo, Spain; E-mail:j.ferrer@cinn.es}}
\footnotetext{\textit{$^{d}$~Institut of Nanotechnology, Karlsruhe Institut of Technology (KIT), D-76344 Eggenstein-Leopoldshafen, Germany}}
\footnotetext{\textit{$^{e}$~Institute de Physique et Chimie de Materiaux de Strasbourg (IPCMS), UMR 7504, CNRS-Universite de Strasbourg, F-67034 Strasbourg, France}}



\sffamily{\textbf{Spin-crossover (SCO) molecules are versatile magnetic switches with applications in molecular electronics and spintronics. Downscaling devices to the single-molecule level remains, however, a challenging task since the switching mechanism in bulk is mediated by cooperative intermolecular interactions. Here, we report on electron transport through individual Fe-SCO molecules coupled to few-layer graphene electrodes \textit{via} $\pi - \pi$ stacking. We observe a distinct bistability in the conductance of the molecule and a careful comparison with density functional theory (DFT) calculations allows to associate the bistability with a SCO-induced orbital reconfiguration of the molecule. We find long spin-state lifetimes that are caused by the specific coordination of the magnetic core and the absence of intermolecular interactions according to our calculations. In contrast with bulk samples, the SCO transition is not triggered by temperature but induced by small perturbations in the molecule at any temperature. We propose plausible mechanisms that could trigger the SCO at the single-molecule level.}}\\


\rmfamily 


Tuning the magnetic properties of individual molecules is sought in molecular spintronics as the key to fabricate switchable molecule-scale electronic components.\cite{Bousseksou2011,Sanvito2011,Lefter2016a} The tuning mechanisms in spin-crossover (SCO) complexes are particularly versatile.\cite{Gutlich2000,Gamez2009,Gutlich2013,SenthilKumar2017} The spin value of the molecular magnetic core, typically an Fe(II) complex, can be switched between a high-spin (HS) and a low-spin (LS) state by modifying its local geometry with light,\cite{Decurtins1985,Hauser1986} pressure,\cite{Gutlich2005,Craig2014} temperature,\cite{Gonzalez-Prieto2011,Dugay2017} voltage\cite{Miyamachi2012, Gopakumar2012,Harzmann2015,Hao2017} or the adsorption of molecules.\cite{Li2010a,Coronado2013,Sanchez-Costa2014} SCO switching with temperature is established in macroscopic crystals where the molecular geometry is well defined and stable in an ordered lattice. The change in spin state can be detected for example as a change in the crystal color\cite{Sanchez-Costa2014} or in the magnetic susceptibility.\cite{Gonzalez-Prieto2011} Additionally, the electrical current has been used as a probe for the spin-state switching in thin films of nanoparticles\cite{Prins2011b,Dugay2017} and molecular thin films.\cite{Lefter2016} 

Downscaling SCO phenomena to the single-molecule level, however, entails a fundamental difference\cite{Miyamachi2012} with respect to measurements on large assemblies: the cooperative intermolecular interactions that mediate the SCO in e.g. crystals\cite{Bedoui2010,Bertoni2016} are absent. To overcome this, alternative strategies based on porphyrin, terpyridine and bispyridine derivatives have recently been developed. The SCO is now induced at the molecular level by either electrostatic effects\cite{Meded2011,Miyamachi2012,Gopakumar2012,Harzmann2015} or molecular stretching\cite{Frisenda2016,Kuang2017} in a single-molecule break-junction. The current through the molecule then acts as the probe to detect the SCO transition where bulk characterization methods fail. At present, it is, however, still unclear what happens for individual molecules that show a temperature-induced SCO transition and are embedded in a solid-state device where the steric hindrance of the crystal felt by the molecule is absent. It is, however, known that structural distortions induced in single molecules by their contacts to metallic (Au) surfaces or electrodes are well known to modify their magnetism\cite{Heersche2006,Gaudenzi2016,Burgess2015,Burzuri2015} and even quench the SCO mechanism.\cite{Warner2013}

Graphene electrodes provide advantages in studying individual temperature-induced SCO molecules. They have proved to be stable from cryogenic up to room temperatures\cite{Burzuri2012b} in contrast to gold nanoelectrodes, enabling the study of temperature-induced SCO transitions in single molecules around room temperature. In addition, a soft molecule-graphene coupling \textit{via} $\pi-\pi$ stacking may contribute to preserve the electronic structure of the molecular orbitals\cite{Garcia-Suarez2013} while providing the flexibility to allow for the SCO transition to occur, as demonstrated for molecules on highly oriented pyrolytic graphite (HOPG) surfaces.\cite{Bernien2015} 

In this work we study electron transport through individual Fe-based SCO molecules linked \textit{via} $\pi-\pi$ stacking to nanometer-spaced few-layer graphene (FLG) electrodes. We observe a reproducible conductance bi-stability between two well-defined states - a strong indication that the spin-crossover switch is active in our single-molecule junctions. In contrast with the temperature-induced transition observed in macroscopic crystals of these molecules, the SCO behavior at the single-molecule level is triggered in time at any temperature even below the well-defined critical temperature for bulk. Density functional theory (DFT) simulations show that the SCO transition induces a sharp change in the energy spectrum of the molecular orbitals leading to conductance bi-stability, in agreement with the experiments. We find that small perturbations (2.5 $\%$) of the distance between the Fe(II) ion and its coordinated ligand atoms can trigger the switch between the HS and LS states. 

\begin{figure}[h]
	\includegraphics{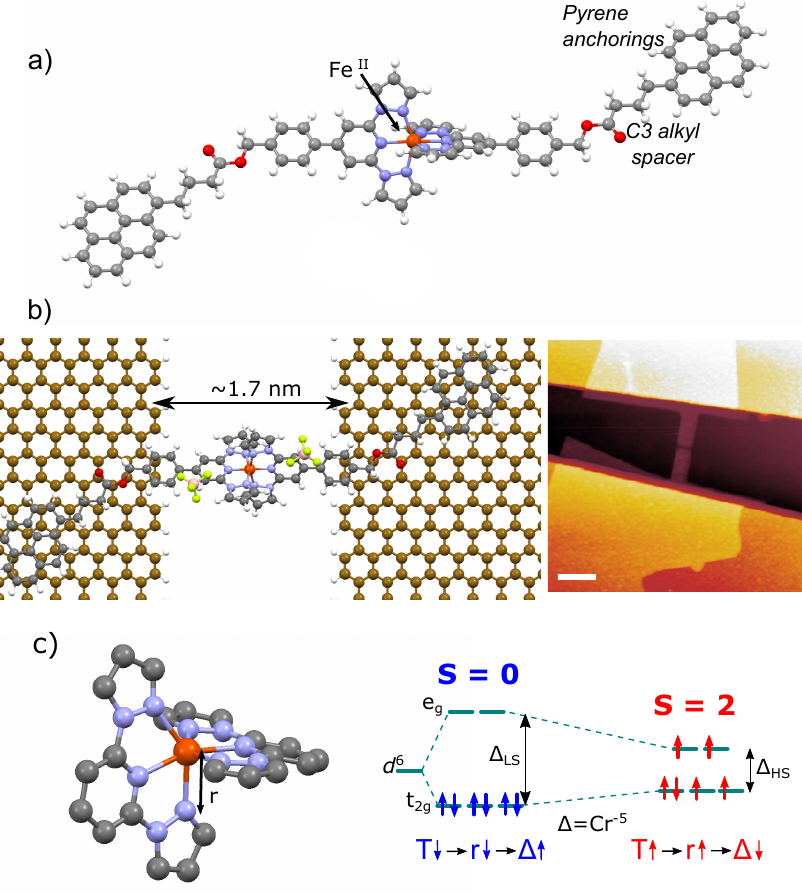}
	\caption{The spin-crossover-FLG single-molecule junction. (a) Molecular structure of the Fe-SCO molecule: an Fe(II) ion is coordinated with two 2,6-bispyrazolylpyridine (bpp) ligands in a distorted octahedral symmetry. Two benzoyl ester and C$_3$ alkyl groups symmetrically connected to the bpp units serve as the backbone of the molecule. Additionally, two pyrene ending-groups provide a soft anchoring to graphene \textit{via} $\pi-\pi$ stacking. BF$_{4}^{-}$ counter anions are omitted for clarity. Color code: C grey, O red, Fe orange, N lilac, F yellow, B pink, H white. The scale bar is 500 nm. (b) Schematics and atomic force microscopy image of a graphene/Fe-SCO/graphene single-molecule junction.  (c) Mechanism of the SCO transition in crystals of Fe-SCO molecules: a change in the Fe-N ligand distance $r$ induced by temperature modifies the crystal field splitting $\Delta$. The different filling of the orbitals leads to a change in the ground state spin from $S$ = 0 to $S$ = 2.}
	\label{fig:figure1}
\end{figure}

We use an [Fe(\textbf{L})$_{2}$](BF$_{4}$)$_{2}\cdot$CH$_{3}$CN$\cdot$H$_2$O molecule,\cite{Gonzalez-Prieto2011} hereafter referred to as Fe-SCO, where \textbf{L} is the ligand 4-(2,6-di(1$H$-pyrazol-1-yl)pyridin-4-yl)benzyl-4-(pyren-1-yl)butanoate. A schematic cartoon of this molecule is shown in Figure \ref{fig:figure1}(a). The Fe(II) ion is coordinated with two 2,6-bispyrazolylpyridine (bpp) ligands in an octahedral (O$_h$) symmetry distorted to a S$_4$ symmetry because of the ligand-Fe(II) coordination specific to this molecule. In addition, our DFT calculations show that the molecule undergoes a Jahn-Teller distortion that reduces the symmetry even further (see Section 3 in the \SI~). The extended backbone is made of two benzoyl ester and C$_3$ alkyl groups symmetrically connected to the bpp units, thus providing extra length and flexibility to the molecule. These ligands are connected to two pyrene ending-groups that promote soft anchoring to graphene \textit{via} $\pi-\pi$ stacking. Additional details on the molecule and its synthesis can be found in Ref.\cite{Gonzalez-Prieto2011}.

The ligand field interaction of approximate O$_h$ symmetry felt by the Fe(II) ion splits the 5-fold degenerate energy spectrum of its 3$d$ electronic shell into two well separated e$_g$ and t$_{2g}$ levels by a ligand field splitting energy $\Delta$  as shown schematically in Figure \ref{fig:figure1}(c). $\Delta$ depends on the average distance $r$ between the  central Fe(II) ion and its neighboring ligand atoms approximately as $r^{-5}$,\cite{Newman2000} and therefore depends inversely on temperature due to thermal expansion. At low temperatures $\Delta$ is larger than the exchange interaction $J$ among the six $d$ electrons in the Fe(II) ion. As a result, the Fe-SCO molecules are in a low-spin (LS) $S$ = 0 ground state (see Figure \ref{fig:figure1}(c)). However, $\Delta$ decreases with increasing temperature and becomes eventually smaller than $J$, making the e$_g$ states accessible. These levels are then filled to maximize the spin according to Hund's first principle, resulting in a high-spin (HS) state $S$ = 2  above a certain temperature (see Figure \ref{fig:figure1}(c)). The magnetic characterization of crystals (bulk) made of these Fe-SCO molecules shows that the spin-crossover transition between HS and LS states occurs at around $T_c$ = 225 K.\cite{Gonzalez-Prieto2011} Interestingly, it is already observed in crystals made of Fe-SCO molecules that structural factors, such as the length or flexibility of the molecule, are determinant to trigger or not the spin-crossover transition.\cite{Gonzalez-Prieto2011}

In our devices, a Fe-SCO molecule is anchored to two FLG electrodes. A schematic representation of the FLG/Fe-SCO/FLG junction and an atomic force microscopy (AFM) image of the graphene electrodes is shown in Figure \ref{fig:figure1}(b). FLG flakes are deposited on a Si/SiO$_2$ substrate by mechanical exfoliation. Thereafter, a nanometer-size gap between source and drain is fabricated using 100-200 nm-wide pre-patterned bridges in the flakes\cite{Island2014} that are narrowed down during electroburning. \cite{Prins2011, Burzuri2012b} See additional details in Section 1.1 of the \SI. Typical gap sizes after electroburning are\cite{Burzuri2016} in the order of 1-2 nm while the length of the molecule spans over 4 nm, so that a substantial part of the molecular backbone may also be lying on the FLG electrodes as depicted in Figure \ref{fig:figure1}(b). The advantage of this configuration is that the anchoring groups lie farther from the edges, facilitating the coupling to graphene by $\pi-\pi$ stacking.\cite{Burzuri2016} In this paper, including the \SI , we discuss six molecular junctions that show molecular features and together they provide a consistent set of data.  

\begin{figure}[h]
	\includegraphics{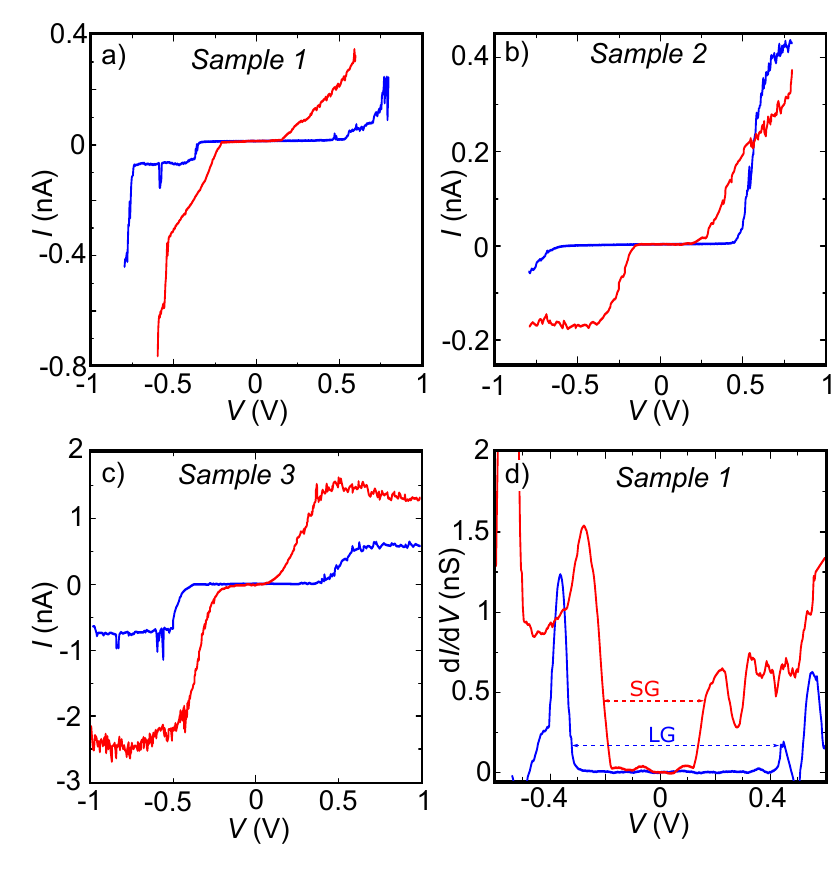}
	\caption{Spin-crossover bistable current-voltage characteristics. (a-c) Current-voltage characteristics measured at $T$ = 4 K in three different junctions containing an Fe-SCO molecule. A clear bi-stability between a small gap (SG, red) and large gap (LG, blue) state is observed. The low-bias current is suppressed in both cases; the blockade is lifted at $V_{\textrm{SG}}\approx 0.1-0.2$ V and $V_{\textrm{LG}}\approx 0.5$ V for the SG and the LG states respectively. The threshold voltages and current levels are approximately reproducible from device to device. (d) Differential conductance d$I$/d$V$ calculated as the numerical derivative of $I$ in sample 1. The change in the size of the low-conductance gap is clearly observed.}
	\label{fig:figure2}
\end{figure}

Figure \ref{fig:figure2}(a-c) shows the current ($I$) - voltage ($V$) characteristics measured at cryogenic temperatures of three different junctions after deposition of the Fe-SCO molecules. When compared with the empty junction, a sharp increase in the current of around two orders of magnitude indicates the formation of a molecular junction (see Section 1.2 in the \SI ). In these three junctions, a clear bi-stability appears between two well-defined states: a large-gap (LG) state (blue curve) and a small-gap (SG) state (red curve). The low-bias current is strongly suppressed in both states. This is a signature of off-resonant transport in a Coulomb blockade regime. At higher bias, sharp resonances at threshold voltages $ V_{\textrm{SG}} \approx$ 0.1-0.2 V in the SG state and $V_{\textrm{LG}}\approx$ 0.5 V in the LG state lift the conductance blockade revealing resonant transport through a molecular orbital with the $V_\textrm{LG,SG}$ being twice the distance to the closest molecular orbital when expressed in energy.\cite{Perrin2015} These blockade and resonant transport features are more clearly seen in the differential conductance d$I$/d$V$ curves shown in Figure \ref{fig:figure2}(d). Sample 5 with a similar bistable behavior is shown in the \SI. An additional sample (sample 4) that does not show a bi-stability appears to be "trapped" in the SG state with similar resonant voltages (see Section 2 in the \SI ). The source of the negative differential resistance observed in samples 3 and 5 could be found in the intrinsic functionalities of some graphene edges' shapes\cite{Carrascal2012} (see Section 2.1 of the \SI .) 

Bistable conductance characteristics have been observed for SCO nanoparticles\cite{Prins2011,Dugay2017} and electrically\cite{Harzmann2015} or mechanically\cite{Frisenda2016} induced SCO molecules coupled to gold electrodes or surfaces.\cite{Jasper-Toennies2017} The sharp change in the resonant voltage from $V_{\textrm{SG}}$ to $V_{\textrm{LG}}$ in our junctions is indicative of an abrupt change in the molecular orbitals around the Fermi level of graphene, tentatively induced by the SCO transition. We further note that current levels and the resonant voltage differ by less than an order of magnitude between the different devices for both the LG and SG states. This reproducibility could stem from the electrode-molecule anchoring geometry facilitated by graphene, as seen in other reports\cite{Ullmann2015,Mol2015,Burzuri2016,Lumetti2016} and predicted in theoretical simulations.\cite{Garcia-Suarez2013}  

\begin{figure}[h]
	\includegraphics{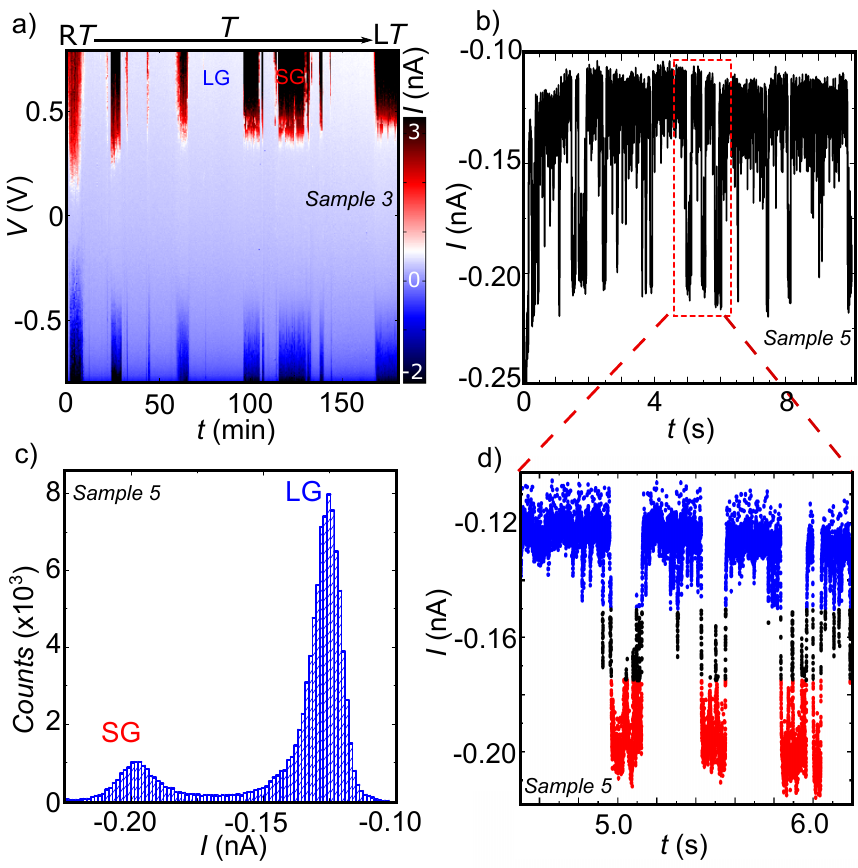}
	\caption{Temperature and time bi-stability. (a) Representative current, $I$, color plot as a function of the bias voltage, $V$, and decreasing temperature measured for sample 3. A telegraph-like switch between two well-defined bistable states occurs independently of temperature. (b) Time-trace of the current measured at a fixed $V$ = -0.6 V and $T$ = 2 K during 10 seconds in sample 5 (see \SI~ for more details of this sample). The same telegraph-like switch between two well-defined bistable states persists well below the bulk SCO transition temperature. (c) Current histogram extracted from (b). The bin size is 1.4$\cdot 10^{-3}$ nA and the total number of counts is $10^{5}$. The LG state centered at -0.126 nA is around an 80$\%$ more stable than the SG state centered at -0.198 nA. The dispersion values taken as the full width half maximum FWHM of the fit to Lorentzian curves are 0.012 nA and 0.017 nA for the LG and the SG respectively. (d) Zoom in of a reduced part of the current trace in (b), showing the two states clearer.}
	\label{fig:figure3}
\end{figure}

Hereafter, we discuss the stability of the SG and LG states as a function of temperature and time. Figure \ref{fig:figure3}(a) shows a representative color plot of $I$ measured as a function of $V$ in sample 3 while decreasing the temperature. A telegraph-like switch is observed between well-separated states. Note, that this kind of time-dependent switch has not been observed in empty graphene junctions nor in junctions containing molecules coupled to graphene \textit{via} $\pi$-$\pi$ but without SCO functionality\cite{Island2014,Mol2015,Ullmann2015,Burzuri2016} and that, as expected in molecular junctions, the current levels are orders of magnitude below those observed in electroburned graphene junctions of carbon chains.\cite{Sarwat2017} In contrast to crystals of these molecules, the switching mechanism at the single-molecule level is triggered in time at any temperature even well below the bulk HS-LS transition temperature. This is \textit{a priori} not surprising since the steric hindrance in the crystal is different than that for a single molecule and the SCO transition may sensitively depend on the local geometry adopted by the molecule in the junction.\cite{Frisenda2016}  

Figure \ref{fig:figure3}(b) shows a current \textit{versus} time trace measured during 10 seconds at a fixed bias voltage of -0.6 V and at 4 K. The voltage is larger than $V_{\textrm{SG}}$ but lower than $V_{\textrm{LG}}$, and $T$ is set well below the SCO transition temperature reported for crystals. The area enclosed in the dotted rectangle is magnified in Figure \ref{fig:figure3}(d). The telegraph-like switch is persistent even at low temperatures. The average lifetime of the states extracted from the plateau lengths is of the order of seconds and tenths of a second for the LG and SG states respectively. These timescales are typical of conformational switches in molecules on surfaces as observed \textit{via} scanning tunnel microscopy\cite{Liljeroth2007,Komeda2011,Zhang2015} and recently, a similar switching in time and timescales of seconds have been observed on SCO molecules deposited on surfaces.\cite{Jasper-Toennies2017} 

Figure \ref{fig:figure3}(c) shows a current histogram obtained from the whole data range in Figure \ref{fig:figure3}(b). Two conductance states clearly emerge above the noise level in the statistics. Interestingly, the LG state centered at -0.126 nA is around 80$\%$ more stable than the SG state centered at -0.198 nA. The fit to a Lorentzian distribution gives dispersion values of 0.012 nA and 0.017 nA for the LG and SG states respectively. The narrow distributions suggest that the molecule switches back-and-forth between two distinct configurations. A similar analysis has been performed on sample 1 in Section 2 of the \SI . Note, that the observed temperature and time dependence of the switching makes charge-offset effects a less likely explanation for it (see Section 1.3 of the \SI ).

To analyze the plausibility of the spin-crossover scenario we have performed density functional theory (DFT) calculations of the structural and electronic properties of the Fe-SCO with the SIESTA code.\cite{Soler2002} Quantum transport simulations of the electrical and spintronic properties of the graphene/Fe-SCO/graphene junctions were carried out with the aid of the code GOLLUM.\cite{Ferrer2014,Garcia-Suarez2012} The simulated molecular junctions contain two graphene electrodes, formed by 282 C and 12 H atoms. The distance between the electrodes is set to 1.7 nm and the Fe-SCO molecule is placed bridging the electrodes so that each pyrene unit lies inside the sheets as indicated in Figure \ref{fig:figure1}(b). The fully optimized molecule-electrode structure results in a $S$ = 0 (LS) ground state. A less stable $S$ = 2 (HS) configuration can also be obtained, which presents longer Fe-N bonds and therefore a smaller ligand field splitting acting on the Fe(II) $d$ orbitals. Additional details can be found in Section 3 of the \SI .

\begin{figure}[h]
	\centering
	\includegraphics{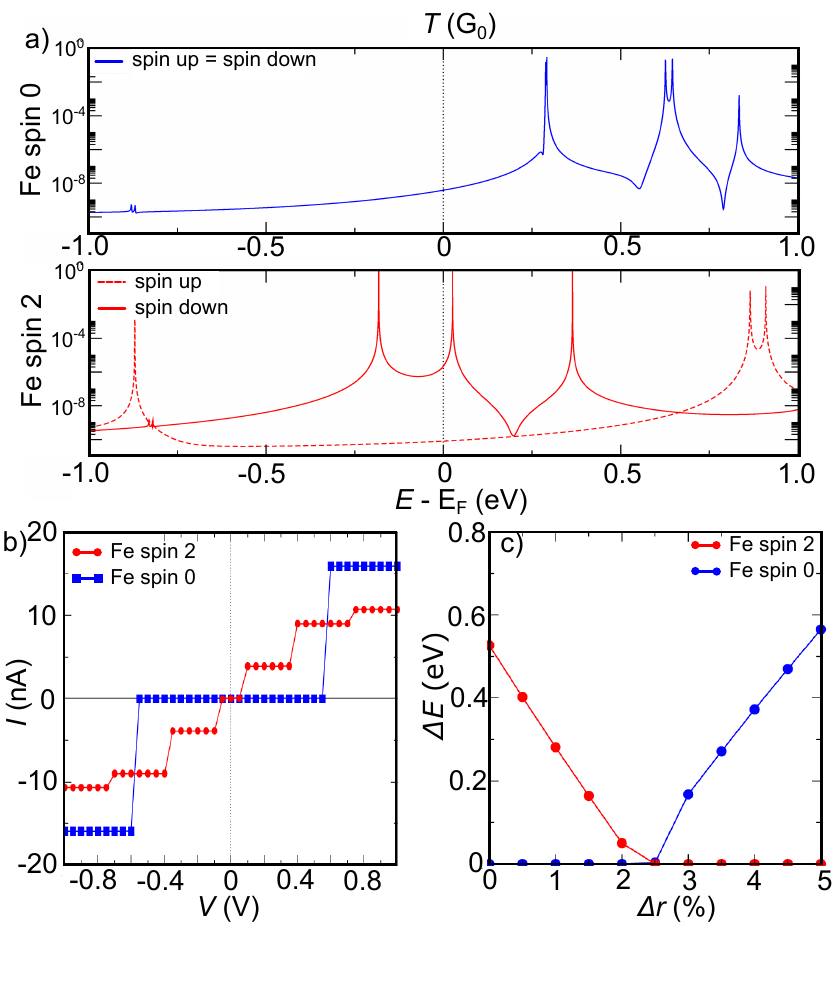}
	\caption{DFT modeling of the SCO transition. (a) Transmission as a function of energy computed for the [top] low-spin (LS) state and the [bottom] high-spin (HS) state. The SCO transition induces a strong modification of the molecular orbitals with the HS LUMO much closer to the Fermi level than the LS LUMO. In addition, the transmission in the HS state is spin resolved up to relatively high energies. (b) Current-voltage characteristics of the LS and HS states calculated from the transmission curves. The change in the level alignment is translated into a change in the current blockade gap as observed experimentally in Figure \ref{fig:figure2}. (c) HS to LS energy difference $\Delta E$ relative to the lowest energy state as a function of the Fe-N bond stretching length $\Delta r$ (see \SI~ for its definition). A small stretching $\Delta r_{c}$ of around 2.5$\%$ can induce the spin-crossover transition form $S = 2$ to $S = 0$. Small perturbations ($< 1 \%$) around that point stabilize LS or HS by more than 0.2 eV.}
	\label{fig:figure4}
\end{figure}

Figure \ref{fig:figure4}(a) shows the spin-dependent transmission function $T_{\sigma}(E)$ computed for the LS and the HS states. The transmission in the LS state is spin degenerate and, at low energies, is mediated by the lowest unoccupied molecular orbital (LUMO) that lies 0.27 eV above the Fermi energy. For the HS state, the structure of $T_{\sigma}(E)$ around the Fermi level is remarkably different. The transmission function depends now on the spin component. The minority spin transmission function $T_\downarrow$ displays three new peaks, one of which (LUMO) is much closer ($\approx$ 0.02 eV) to the Fermi level than in the LS state. This shift has a strong effect on the computed $I$-$V$ characteristics for the HS and the LS states as shown in Figure \ref{fig:figure4}(b). The calculated curves strikingly resemble to the experimental ones in Figure \ref{fig:figure2} in terms of resonant voltages ($V_{\textrm{LG}}$ and $V_{\textrm{SG}}$) and current levels. By comparison, we can thus ascribe the SG state in the measurements to the HS state and the LG state to the LS state. This correspondence is in agreement with previous experimental reports with other SCO molecules.\cite{Harzmann2015,Frisenda2016}

In contrast with $T_\downarrow(E)$, $T_\uparrow(E)$ is rather flat and smaller in a wide energy range. We believe that this feature could be exploited to fabricate switchable spin filters. However, for that to happen the spin direction on the molecule itself needs to be fixed for instance by introducing magnetic anisotropy. The explanation behind this spin-resolved transmission can be attributed to the different filling of the orbitals as explained in more detail in Section 3 of the \SI . 

We have further analyzed the impact of the molecule's stretching on the relative energies of its HS and LS states. However, DFT may not provide an accurate enough description of the multi-electronic state of the molecule, so we have fitted the DFT estimate of the ligand field effects for the LS state to the following multi-electronic Hamiltonian:

\begin{equation}
H=H_{EE}+H_{LF},
\end{equation}

\noindent
where $H_{EE}$ represents the electrostatic interaction between the $d$ electrons of the Fe(II) ion and $H_{LF}$ accounts for the ligand field interaction of the $d$ electrons with the surrounding atoms. $H_{EE}$ can be parametrized in terms of the Slater-Condon parameters $F_k$.\cite{Griffith1971} We have taken $r_0$ as the average distance between the central Fe(II) ion and the surrounding ligands in the most stable configuration of the molecule, and have then stretched this average distance by an amount $\Delta r$. Because the strength of the ligand field is known to be approximately proportional to $r^{-5}$ for 3$d$ electrons,\cite{Newman2000} we have simulated the molecule's stretching by decreasing the strength of the $H_{LF}$ term by $\Delta r^{-5}$. The total energies of the lowest LS and HS multi-electronic solutions, relative to the lowest state, are plotted in Figure \ref{fig:figure4}(c). Note, that the spin transition from $S$ = 0 to $S$ = 2 occurs at a $\Delta r_{c}$ as small as 2.5 $\%$. Importantly, we note that variations of $\pm$1 $\%$ relative to $\Delta r_{c}$ stabilize either of the two states by more than 0.2 eV. Translating this effect to single-molecule junctions, we expect that small perturbations to the molecule arrangement inside the gap because of bias, temperature fluctuations, vibrations, etc may trigger the spin-switching behavior even below the bulk $T_c$, or hinder it (as in sample 4 in the Supporting Information) depending on the junction conformation. Note, that the transmission curves shown in Figure \ref{fig:figure4}(a) for the LS (HS) are representative of any $\Delta r$ below (above) 2.5 $\%$ with minor variations.

To further study the effect of perturbations we have displaced in the simulations the molecule perpendicular to the junction gap. We have found that the most stable configuration corresponds to the molecule's core lying inside the gap. We have also found that the molecule may drift easily inside the gap with energy barriers as small as 20 meV. However, once the core hits one of the two electrodes, the energy barrier for further motion above the electrode rises sharply to 0.5 - 1 eV because of steric hindrance. As a consequence, the molecule's core stays within the gap. Importantly, the relative stability of the different magnetic solutions of the Fe-SCO is not affected by the details of the bonding to the sheets or the orientation.

A final remark concerns the spin-state life times and the associated stimuli needed to switch the SCO molecule. Long spin-state lifetimes at room temperature have been reported for SCO complexes in solution and on surfaces, where cooperative interactions and low-energy phonons proper of a SCO crystal are absent.\cite{Stock2013,Stock2016,Jasper-Toennies2017} These spin transitions have been observed to involve large structural rearrangements leading to high energy barriers between the LS and HS states. Similar structural distortions have been observed for bpp derivatives.\cite{KershawCook2015,Zhang2015a} Slow, temperature-independent SCO up to 120 K has been found in other Fe(II) complexes that share the same ligand geometry as bpp molecules.\cite{Renz2000,Hauser2006} These two ingredients: lack of cooperative intermolecular interactions and a bpp skeleton are present in our solid-state device. The calculations (see Section 3 of the \SI) for the Fe(II) SCO molecule in this study indicate that tunneling is the dominant switching mechanism up to temperatures of the order 100 K in the absence of stimuli other than temperature, in agreement with Ref.\cite{Xie1987}. We find resident times in the range 10$^{-3}$ - 10$^0$ seconds independent of the temperature (see Section 3 of the \SI), that are consistent with our experiments. For higher temperatures vibrational heating reduces those resident times. We note here that, the temperature of a single-molecule that is voltage biased through weakly coupled electrodes is ill-defined and may be subject to fluctuations.\cite{Ward2011} 

Additionally, other stimuli, i.e., mechanical, electrical or thermal, could contribute to triggering the SCO mechanism or block it if a single-molecule is embedded in a junction. Transitions between HS and LS states can be induced by the electronic charging of the molecule by an external bias.\cite{Hao2017} Local perturbations to the binding geometry could, for example, induce strain to the molecule ligands, which can be large enough to stretch the ligand-metal ion distance to induce the spin transition.\cite{Kuang2017} Moreover, the strong electric fields generated by the voltage in these narrow solid-state nano-junctions could induce non-equilibrium spin-orbit effects\cite{Meded2011} or dipole-induced strain\cite{Weston2015} in the molecule due to the opposite charge of the Fe core and the ligands. Again, these effects can influence the ligand-metal ion distance and thereby initiate the SCO transition.

In conclusion, we have measured electron transport through individual Fe-SCO molecules coupled to few-layer graphene electrodes. We observe a reproducible switching between two bistable states triggered by the SCO transition in the molecule. DFT calculations provide a qualitative and quantitative agreement thanks to the well-defined geometry of the molecule-graphene coupling. The switching occurs well below the critical temperature for crystals of the same molecules. DFT suggests that the switch at the single-molecule level can be induced by small perturbations to the ligand distance in the molecular junction. Finally, the HS state of the molecule is spin resolved; these molecules could therefore be used as switchable spin polarizers.

\section*{Conflicts of interest}
There are no conflicts to declare.

\section*{Acknowledgments}

We acknowledge financial support from the Dutch Organization for Fundamental research (NWO/FOM), the European Commission through an advanced ERC grant (Mols@Mols) and the Marie Curie ITN MOLESCO, from the Netherlands Organization for Scientific Research (NWO/OCW) as part of the Frontiers of Nanoscience program, and from the Spanish Ministerio de Economia y Competitividad through the project FIS2015-63918-R. EB thanks funds from the EU FP7 program through Project 618082 ACMOL and NWO through a VENI fellowship.


\clearpage
\onecolumn	

{\noindent\LARGE{\textbf{ Supporting Infromation for Spin-State dependent Conductance Switching in Single Molecule-Graphene Junctions$^\dag$}} }
\vspace{0.6cm} \\

	This supporting information is divided in three sections. Section 1 contains additional details on the geometry and dimensions of the electrodes and a representative electronic characterization of the empty few-layer graphene (FLG) electrodes. Section 2 contains additional examples of FLG$|$Fe-SCO$|$FLG junctions. Finally, Section 3 contains more details of the DFT and quantum transport simulations, including the projection of the density of states on the different molecular orbitals.

	\section{Geometry and characterization of empty FLG junctions}
	
	\subsection{Geometry and dimensions of the FLG devices. Molecular deposition.}
	
\begin{figure}[b]
		\centering
		\includegraphics[width=0.7\textwidth]{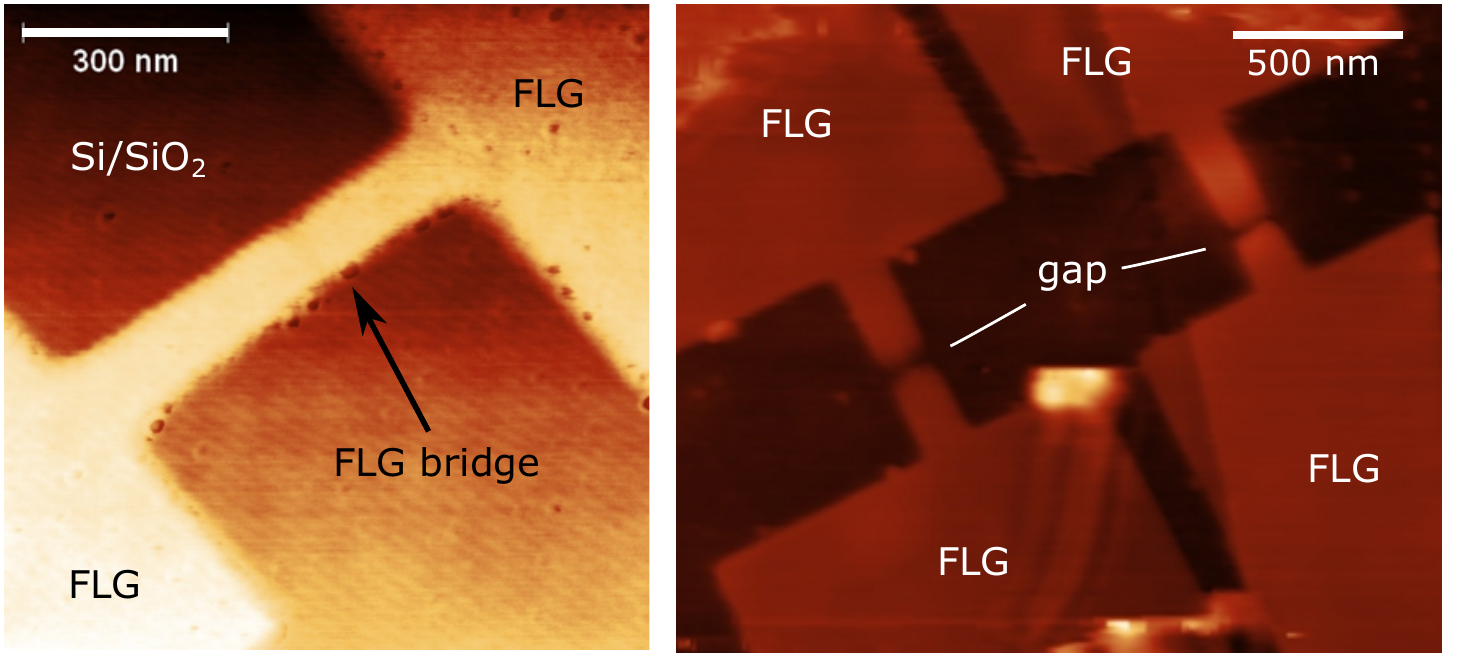}
		\caption{Atomic force microscopy (AFM) images of several FLG devices after pre-patterning with a combination of electron-beam lithography and oxygen plasma etching.}
		\label{FigureS11}
	\end{figure}

	A thorough description of the fabrication of the FLG electrodes as well as a set of images of the devices can be found in Ref.\cite{Island2014} . In summary, a micron-size flake is selected under the optical microscope to fabricate the FLG electrodes. Gold pads are deposited thereafter on the flake to create electrical access. Parts of the flake are etched away with oxygen plasma so that a pre-patterned bridge of approximately 100-200 nm width is formed as seen in Figure 1 of the main manuscript. Additional atomic force microscopy (AFM) images are shown in Figure \ref{FigureS11}. The gap separating source and drain electrodes is opened by feedback-controlled electroburning. The burning process typically starts around the middle segment of the bridges' edges as explained in Ref.\cite{Island2014} . The minimum separation between the electrodes, 1-2 nm, is localized in a small section of the gap where the last contact point before total electroburning was placed. The remaining sections of the electrodes edges can be separated tens of nanometers apart.

	Once the electrodes are prepared, the SCO molecules are dispersed in a 10$^{-4}$ M solution of acetonitrile. A droplet is drop-casted on the chip and thereafter dried with N$_2$ gas. A low concentration is chosen on purpose to reduce the probability that a given device has its electrodes bridged by more than one molecule. Furthermore, only those molecules that are deposited at the effective gap section can bridge the electrodes. 
	
	We note that only 10$\%$ of the samples contain molecular traces, so the probability that one molecule bridges the electrodes is $P_1\approx 0.1$. This value is comparable to other studies using electromigrated Au\cite{Perrin2015}, or electroburned graphene electrodes\cite{Burzuri2016}. This spells that the probability of finding two molecules bridging a junction is  $P_2=P_1^2\approx 0.01$, if the two events are statistically uncorrelated, so that only one in one hundred samples should have more than one molecule. In addition, steric hindrance among molecules should reduce $P_2$ below 0.01. As a consequence, we expect that the probability that one of our samples contain more than one molecule is  low.
	
	In the unlikely scenario of more than one molecule in the gap and bridging the electrodes, the electron-transport across the junction would be dominated by the one more strongly coupled to the electrodes. In the even more unlikely circumstance that two molecules are present with equivalent couplings, three different current values should be observed for switching molecules (2$I_{min}$, $I_{min}$+$I_{max}$, 2$I_{max}$) given that the molecules switch incoherently. The latter is ruled out in our experiments. Our conclusions remain trivially the same if they switch coherently.

	\subsection{Characterization of the empty gap}
	
\begin{figure}
		\centering
		\includegraphics{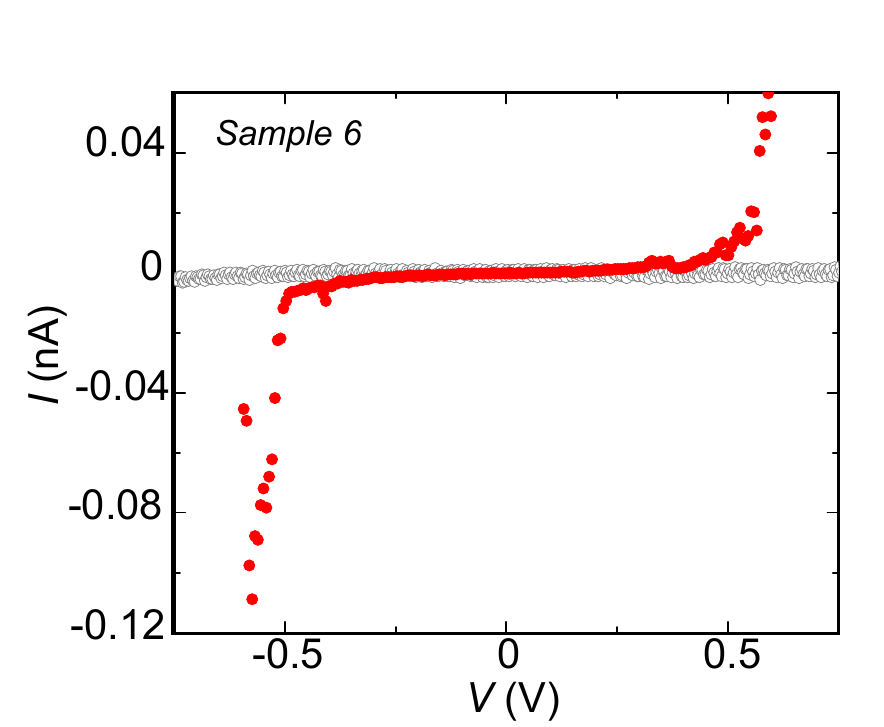}
		\caption{Current versus bias voltage measured across a pair of nanometer-spaced FLG electrodes (sample 6) before (gray empty dots) and after (red full dots) deposition of the Fe-SCO molecules. The high-bias current increases around two orders of magnitude indicating the formation of a molecular junction.}
		\label{FigureS1}
	\end{figure}

	Figure \ref{FigureS1} shows the current ($I$) - voltage ($V$) characteristic measured between a pair of FLG electrodes (sample 6) before (empty electrodes) and after deposition of a solution with the Fe-SCO molecules. The high-bias current increases around two orders of magnitude after molecular deposition pointing to the formation of a molecular junction\cite{Island2014,Burzuri2016}. The low-bias conductance gap is consistent with the large-gap (LG) state observed in the other samples and with the Fe-SCO molecule in a LS state according to DFT calculations.
	
	\subsection{Charge offset effects}
	
	Charge-offsets can lead to abrupt shifts in the current but these are, however, typically "seen in" and "activated by" sweeping the gate voltage. In contrast, our measurements show time-dependent switching at a fixed gate voltage. Furthermore, the switching rate associated with thermally-activated hoping of charges from charge trap to charge trap would be strongly temperature dependent. In contrast, we do not observe a strong temperature dependence of the switching in Figure 3(a) in the main manuscript. Deep traps may still be present and activated by an electrical field. However, the randomness we see in the switching rates (both in time and bias voltage) together with the reproducibility of the threshold voltages mentioned before do no support such a picture.
	
	\section{Additional samples}

\begin{figure}
		\centering
		\includegraphics[width=0.7\textwidth]{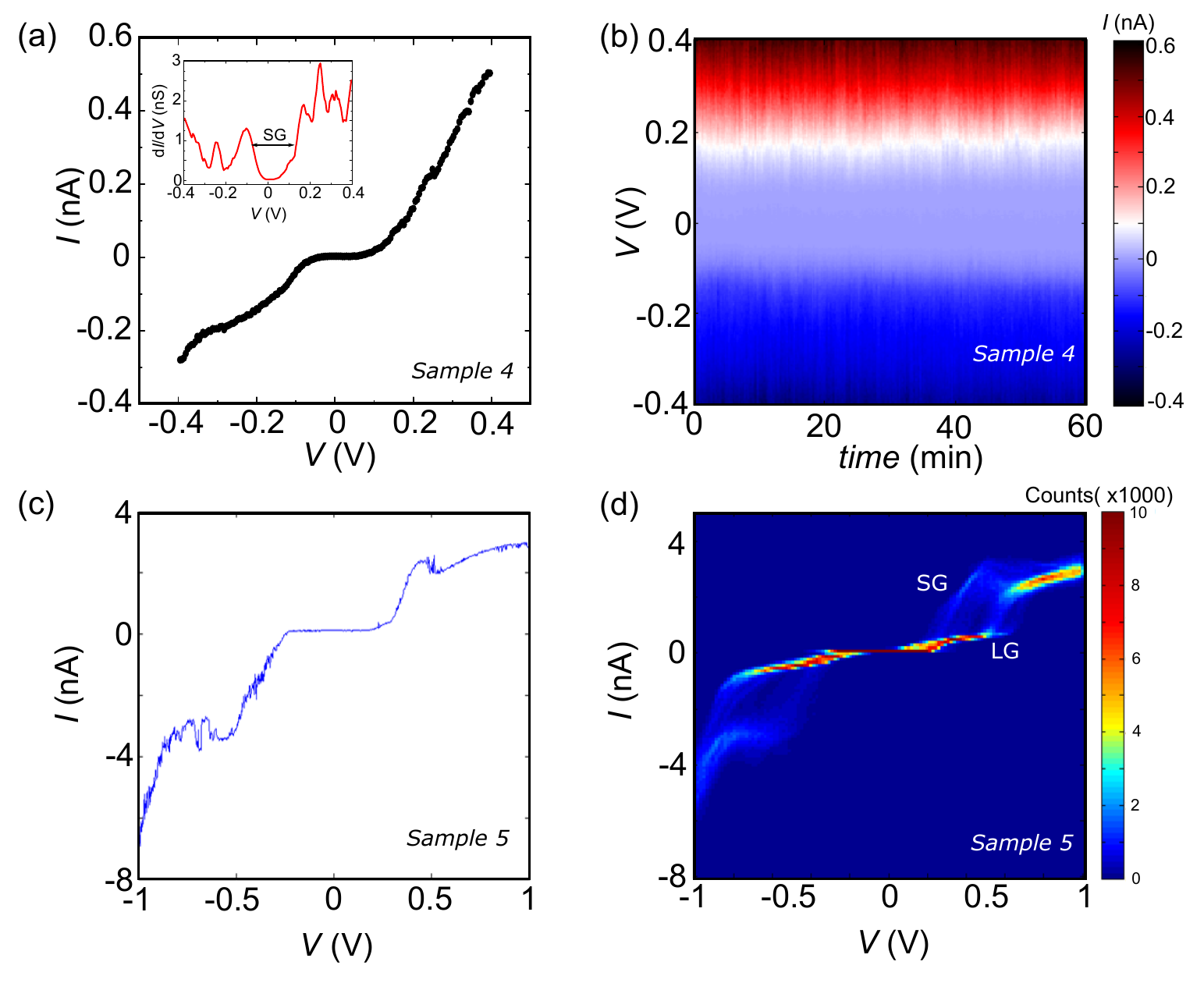}
		\caption{(a) $I$ versus $V$ measured in sample 4 at $T$ = 100 K. The inset shows the differential conductance d$I$/d$V$ numerically derived from $I$. The conductance gap is consistent with a SG state. (b) Color plot of $I$ measured as a function of $V$ and time. No conductance bistability or switching to a LG state is observed. (c) $I$ versus $V$ measured in sample 5. Negative differential conductance is observed symmetric in bias. (d) $I$ versus $V$ histogram made of $10^4$ traces measured over time in sample 5. Two bistable conductance states can be observed.}
		\label{FigureS21}
	\end{figure}

	Figure \ref{FigureS21}(a) shows an $I$ - $V$ characteristic measured at $T$ = 100 K in an additional junction (sample 4) containing an Fe-SCO molecule. The high-bias current and the size of the low-bias suppression gap fit well with the SG state characteristics of samples 1, 2 and 3 shown in Figure 2 of the main manuscript. Following the rationale provided by DFT, the Fe-SCO molecule is in a high-spin (HS) state as explained in Figure 4(b) of the main manuscript. In contrast to samples 1 to 3, no conductance bistability or switching between different conductance states is observed in this junction. The time stability of this sample is plotted in Figure \ref{FigureS21}(b), showing that the molecule is locked in the small-gap state. This different behavior could be explained for a molecule that undergoes relatively larger deformations between the electrodes ($\Delta r > 2.5 \%$). Small perturbations can in that case not modify the ligand distance enough to induce the spin crossover transition (see Figure 4(c) in the main manuscript).

	Figure \ref{FigureS21}(c) shows an $I$-$V$ characteristic measured on sample 5 of which the time trace is shown in Figure 3 of the main text. The low-bias conductance gap is consistent with the SG state and therefore with the HS state. The high-bias current shows negative differential conductance. We discuss this in the next subsection. Figure \ref{FigureS21}(d) shows a 2D $I$ \textit{versus} $V$  histogram made with approximately $10^4$ current traces measured over time. Two bistable conductance states, corresponding to the LG and the SG states can be observed consistent with samples 1 to 3 in the main manuscript. The current level dropped around one order of magnitude between these measurements and the time trace shown in Figure 3 of the main manuscript, possibly due to a rearrangement of the molecule in the junction. Importantly, the bistable character of the molecular junction was preserved.

	\begin{figure}
		\centering
		\includegraphics[width=0.7\textwidth]{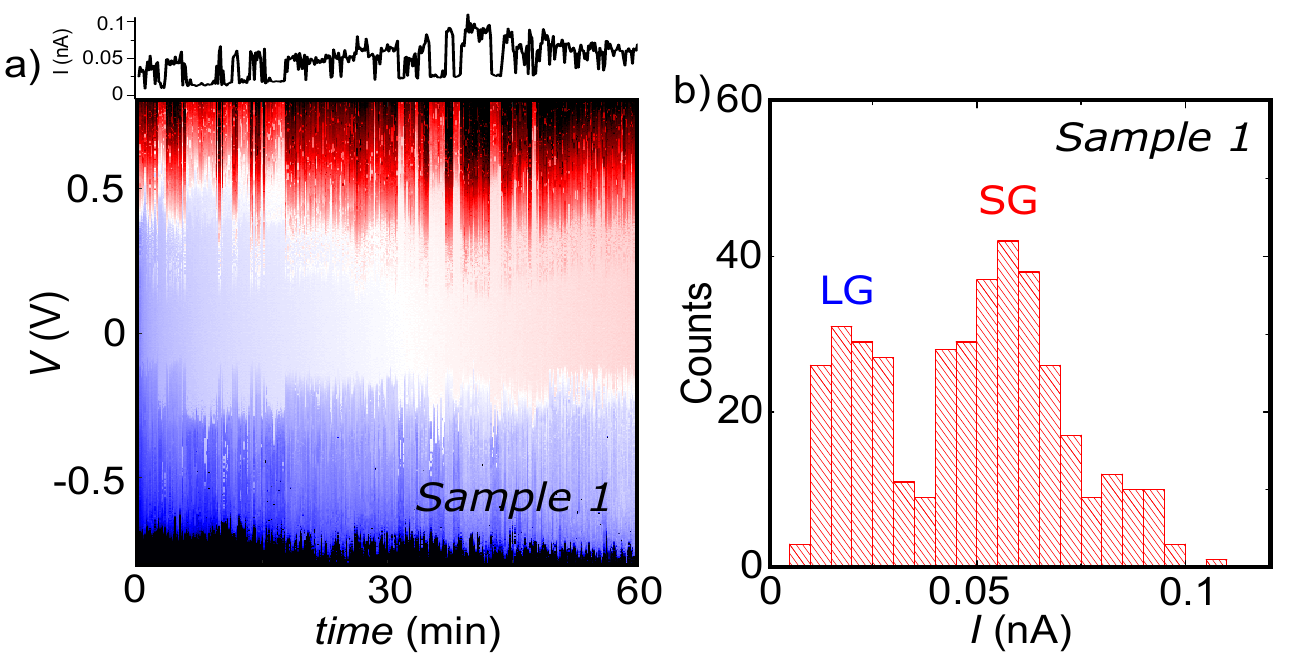}
		\caption{(a) Representative current color plot as a function of bias voltage and time measured for sample 1. The top current trace is taken from the color plot at $V$ = 0.48 V. (b) Current histogram extracted from (a) at $V$ = 0.48 V. The bin size is 0.005 nA and the total number of counts is 401.}
		\label{FigureS22}
	\end{figure}

	Figure \ref{FigureS22}(a) shows the time dependence of the $I$-$V$ characteristics of sample 1 shown in the main manuscript. A switching in time between two states, similar to the one shown in Figure 3 of the main manuscript, is observed. This is more clearly seen in the single $I$ trace taken at $V$ = 0.48 V from the plot (top of Figure \ref{FigureS22}(a)). A histogram of the current at that bias voltage is shown in Figure \ref{FigureS22}(b). Two conductance values denoted as LG and SG emerge from the noise. The high-current tails of both peaks is more extended in $I$, probably due to a drift of the global current to higher values as already observed in the $I$ trace in Figure \ref{FigureS22}(a).
	
	\subsection{Negative differential resistance}
	
	Sample 3 shows clear negative differential resistance (NDR) at positive bias. In addition, sample 5 shows symmetric NDR at positive and negative bias, as seen in Figure \ref{FigureS21}(c). This behavior has been reported for SCO molecules between gold electrodes\cite{Harzmann2015}. NDR in a molecular junction is a complex phenomenon where several mechanisms could be underlying: vibrational modes, interfering transport paths within the molecule and different electrode-molecule couplings among others, see Ref.\cite{Perrin2014} and references therein for a detailed description. The transmission functions shown in Figure 4 of the main manuscript seem to discard the interference scenario since no suppression of the transmission is observed. 
	
	Additionally, in the  case of graphene electrodes, it is predicted that sharp zig-zag edges could yield NDR because they carry localized edge states if contacted appropriately; other edge configurations convey trivial functionalities\cite{Carrascal2012}. Interestingly, the molecular features in the conductance are predicted to be universal for graphene electrodes, in the sense that do not depend on the details of the molecule-electrode contact. This effect is due to the physisorbed nature of our contacts and has been discussed in detail in Ref.\cite{Garcia-Suarez2013} . The atomic-scale shape of our graphene edges varies from device to device and therefore we expect some variability in the conductance features (like NDR) superimposed to the reproducible features stemming from the molecule (the bi-stable switch).

	\section{DFT calculations}
	\subsection{Density of states in the LS and HS states}

\begin{figure}
		\centering
		\includegraphics[width=0.7\textwidth]{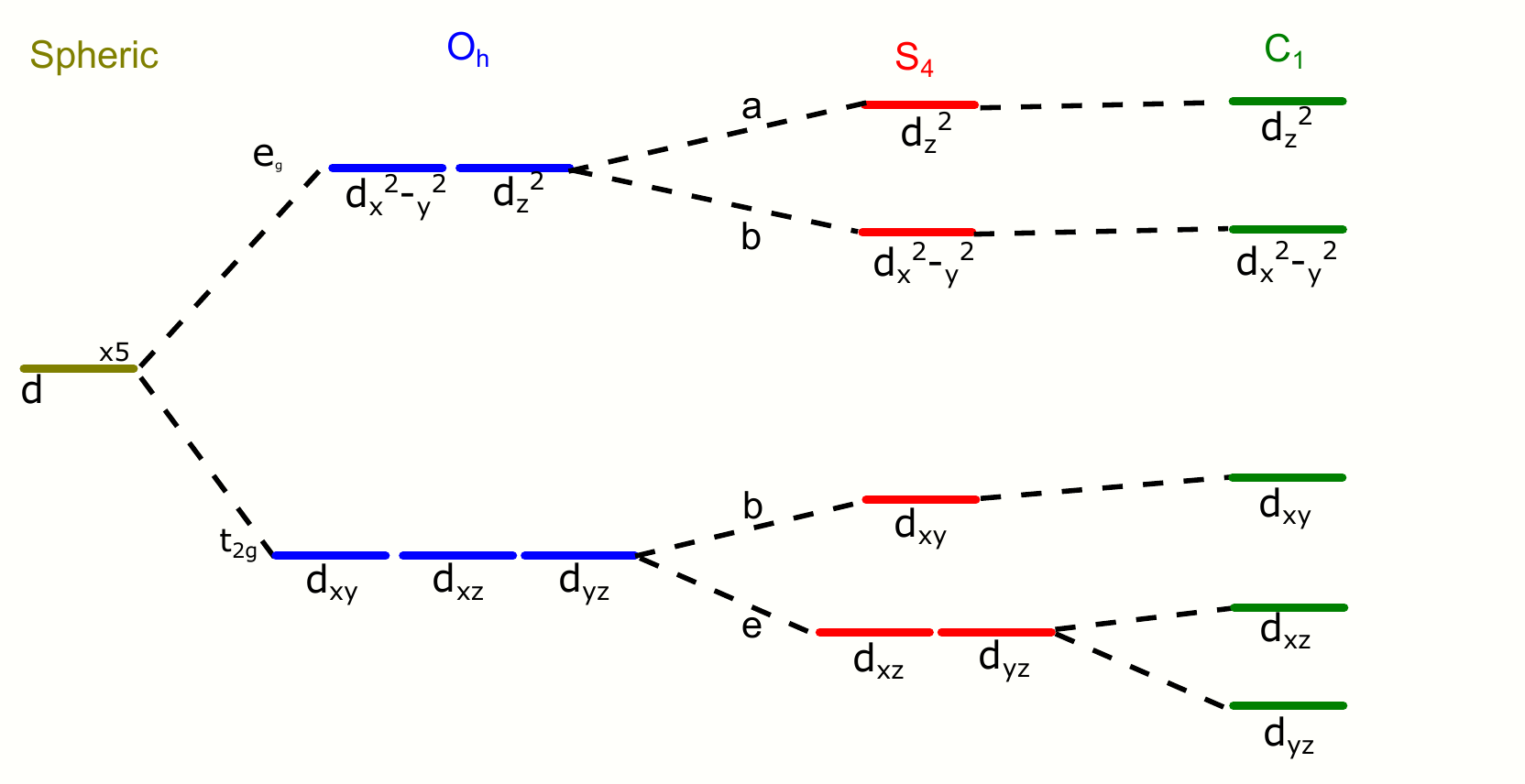}
		\caption{Components of a metallic d orbital splitting under octahedral (O$_h$), tetrahedral (S$_4$) and C$_1$ symmetry.}
		\label{FigureS30}
	\end{figure}

	Figure \ref{FigureS30} shows the effect of the molecular ligand field on the metallic $d$ orbitals of the Fe(II) atom. The five $d$ orbitals initially degenerated in the free ion are split into e$_g$ and t$_{2g}$ states by an octahedral symmetry. A further reduction of the ligand symmetry to S$_4$ splits further the levels.

	\begin{figure}
		\centering
		\includegraphics[width=1\textwidth]{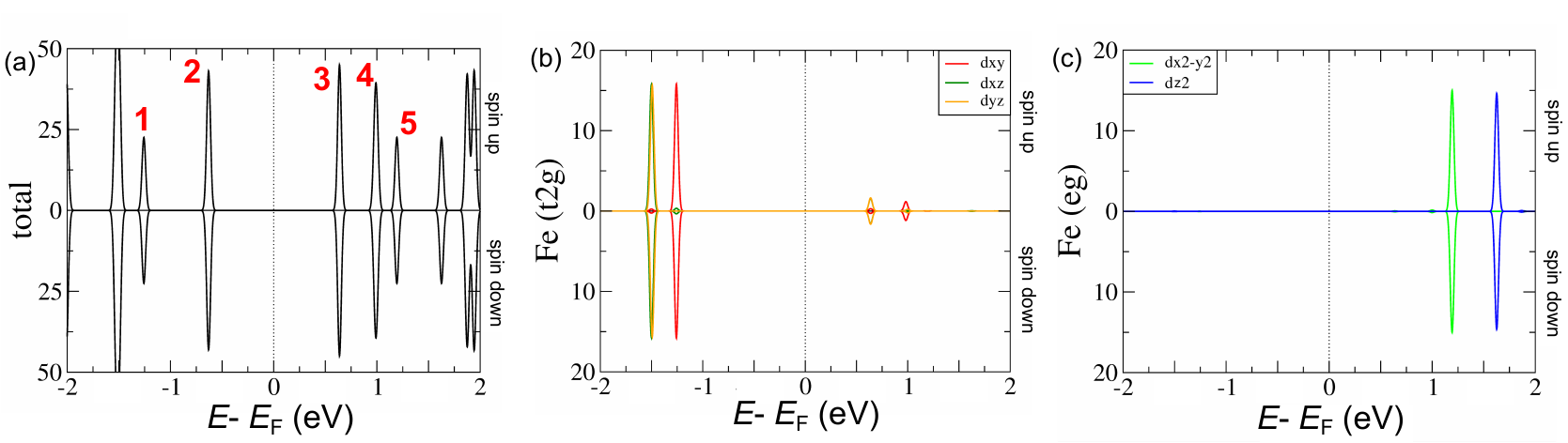}
		\caption{(a) Total projected density of states for the isolated Fe-SCO molecule in the LS state. The red numbers correspond to the states plotted in Figure \ref{FigureS32}. (b) Local projected density of states associated to the t$_{2g}$ orbitals of Fe in the LS state. (c) Local projected density of states associated to the e$_g$ orbitals of Fe in the LS state.}
		\label{FigureS31}
	\end{figure}
	
	\begin{figure}
		\centering
		\includegraphics[width=0.5\textwidth]{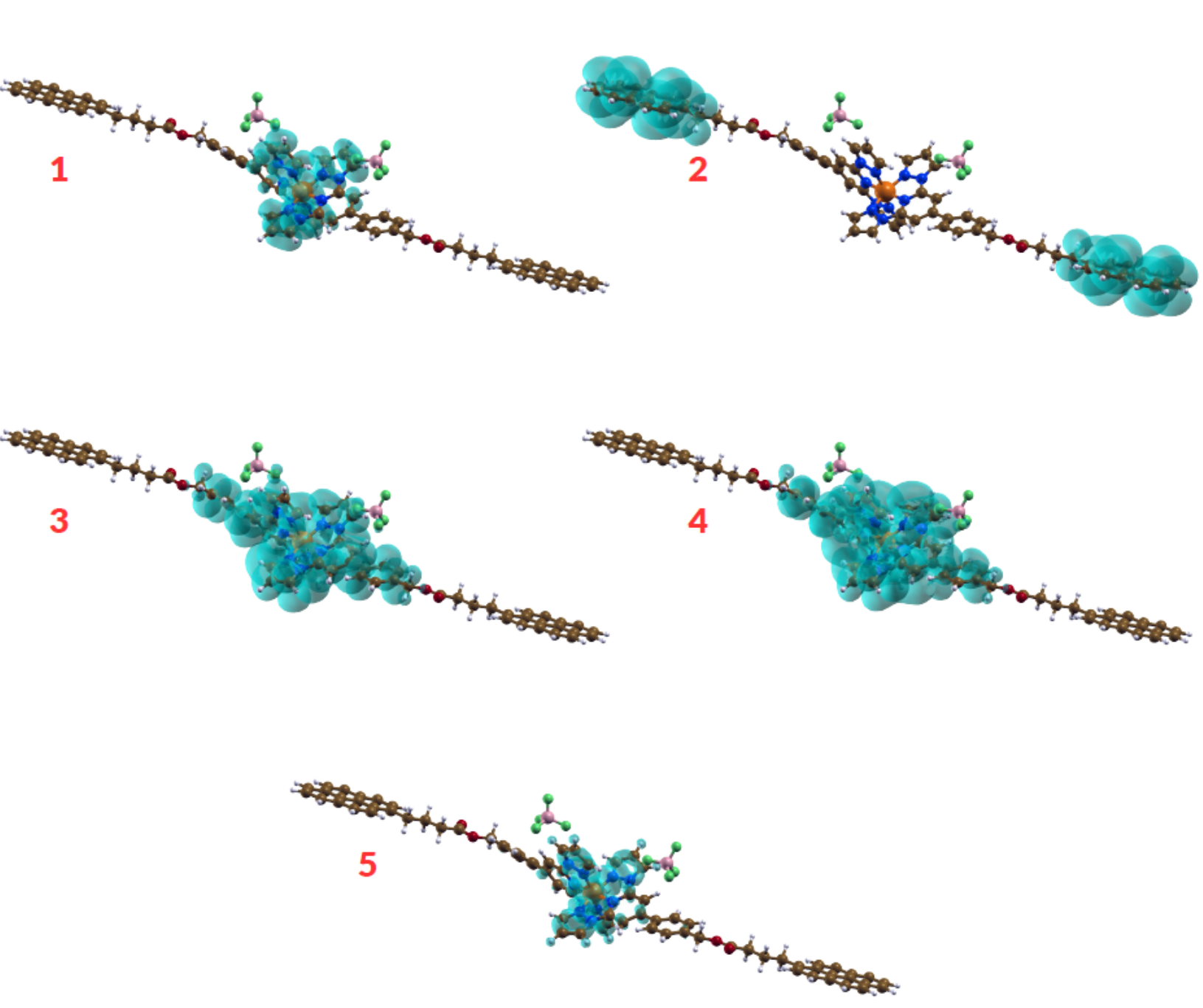}
		\caption{Local density of states of the five orbitals closer to the Fermi level for the Fe-SCO molecule in the LS state. The numbers correspond to those in Figure \ref{FigureS31}. Note that the conjugation is broken at the oxygen atoms; electron transmission through the molecule is therefore most likely facilitated by electron injection through the molecular backbone, instead of the pyrene groups as previously reported\cite{Burzuri2016}.}
		\label{FigureS32}
	\end{figure}
	
	The calculated projected density of states of the isolated Fe-SCO in the LS state is shown in Figure \ref{FigureS31}(a). As explained in the main manuscript the density of states is spin degenerate. The d$_{xy}$, d$_{yz}$ and d$_{xz}$ states (t$_{2g}$) lie mainly below the Fermi level while the d$_{x^{2}-y^{2}}$ and d$_{z^{2}}$ states (e$_g$) are found well above as seen in the projected density of states in Figure \ref{FigureS31}(b) and Figure \ref{FigureS31}(c) associated to the t$_{2g}$ and e$_g$ states respectively. The local density of states of the orbitals closer to the Fermi level is plotted in Figure \ref{FigureS32}. The states 3 and 4, with small contributions of the t$_{2g}$ orbitals, are strongly delocalized over the Fe-SCO core up to the oxygen atoms where the conjugation is broken. The states 1 and 5, mainly associated with the d$_{xy}$ and d$_{x^{2}-y^{2}}$ orbitals, which are transversal to the direction of the molecule, are slightly delocalized. The main contribution of d$_{yz}$ and d$_{xz}$ is found well below the Fermi level and d$_{z^{2}}$ is at higher energies. The state 2 is mainly delocalized over the pyrene anchoring groups.

	\begin{figure}
		\centering
		\includegraphics[width=1\textwidth]{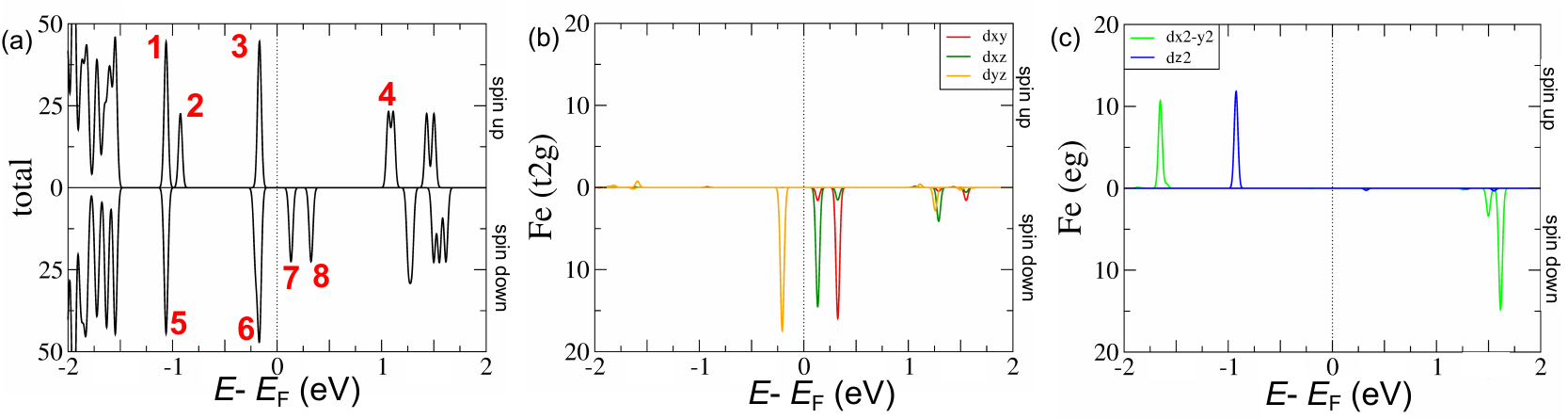}
		\caption{(a) Total projected density of states for the isolated Fe-SCO molecule in the HS state. The red numbers correspond to the states plotted in Figure \ref{FigureS34}. (b) Projected density of states associated to the t$_{2g}$ levels of Fe in the HS state. (c) Projected density of states associated to the e$_{g}$ levels of Fe in the HS state.}
		\label{FigureS33}
	\end{figure}

	\begin{figure}
		\centering
		\includegraphics[width=0.7\textwidth]{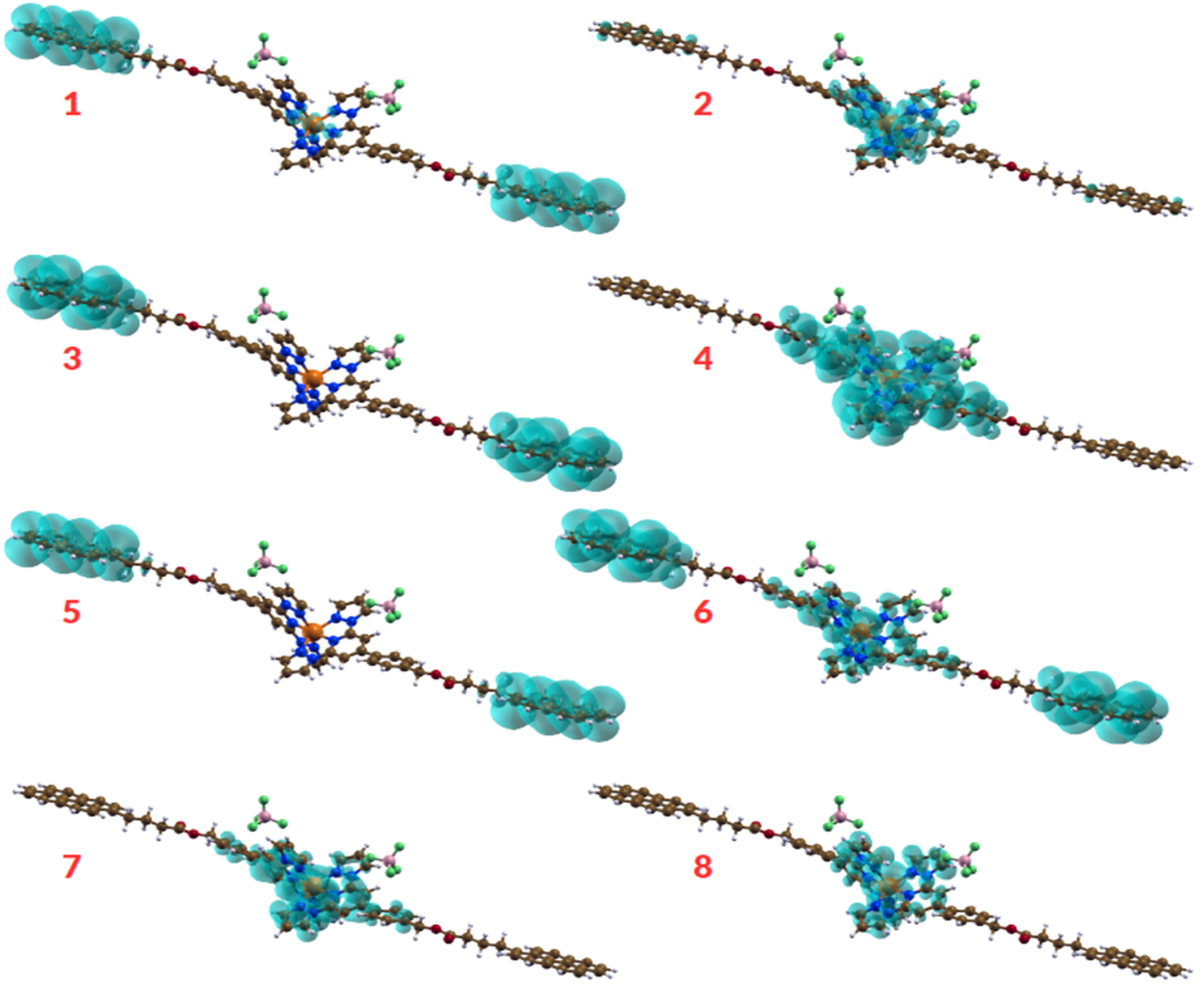}
		\caption{Local density of states of the seven orbitals closer to the Fermi level for the Fe-SCO molecule in the HS state. The numbers correspond to those in Figure \ref{FigureS33}}
		\label{FigureS34}
	\end{figure}

	The magnetic behavior of the Fe-SCO in the HS state is rather different. The calculated projected density of states of the isolated Fe-SCO in the HS sate is shown in Figure \ref{FigureS33}(a). The density of states is now spin resolved. Both the t$_{2g}$ and e$_g$ orbitals are occupied and well below the Fermi level for the majority spin component, while only d$_{yz}$ from the t$_{2g}$ orbitals is occupied for the minority spin component (see Figure \ref{FigureS33}(b) and \ref{FigureS33}(c)). The minority spin t$_{2g}$ states are only partially occupied, and this fact leads them to be placed quite close to  the Fermi level. Indeed, as the symmetry of the Fe-SCO is reduced from O$_h$ to S$_4$, the e$_g$ states split into a and b states, and the t$_{2g}$ split into b and e states as seen in Figure \ref{FigureS30}. The doubly-degenerate states are further split by a Jahn-Teller distortion. Because these turn out to be the two states lying closest to the Fermi energy, this final splitting has a strong impact on $T_\downarrow$ around the Fermi level. As a result of the orbital filling, the spin up transmission is mainly blocked up to high energies and only the spin down is transmitted, as explained in the main manuscript. The molecule could therefore be used as a spin filter up to relatively high energies.
	
	\begin{figure}
		\centering
		\includegraphics[width=0.7\textwidth]{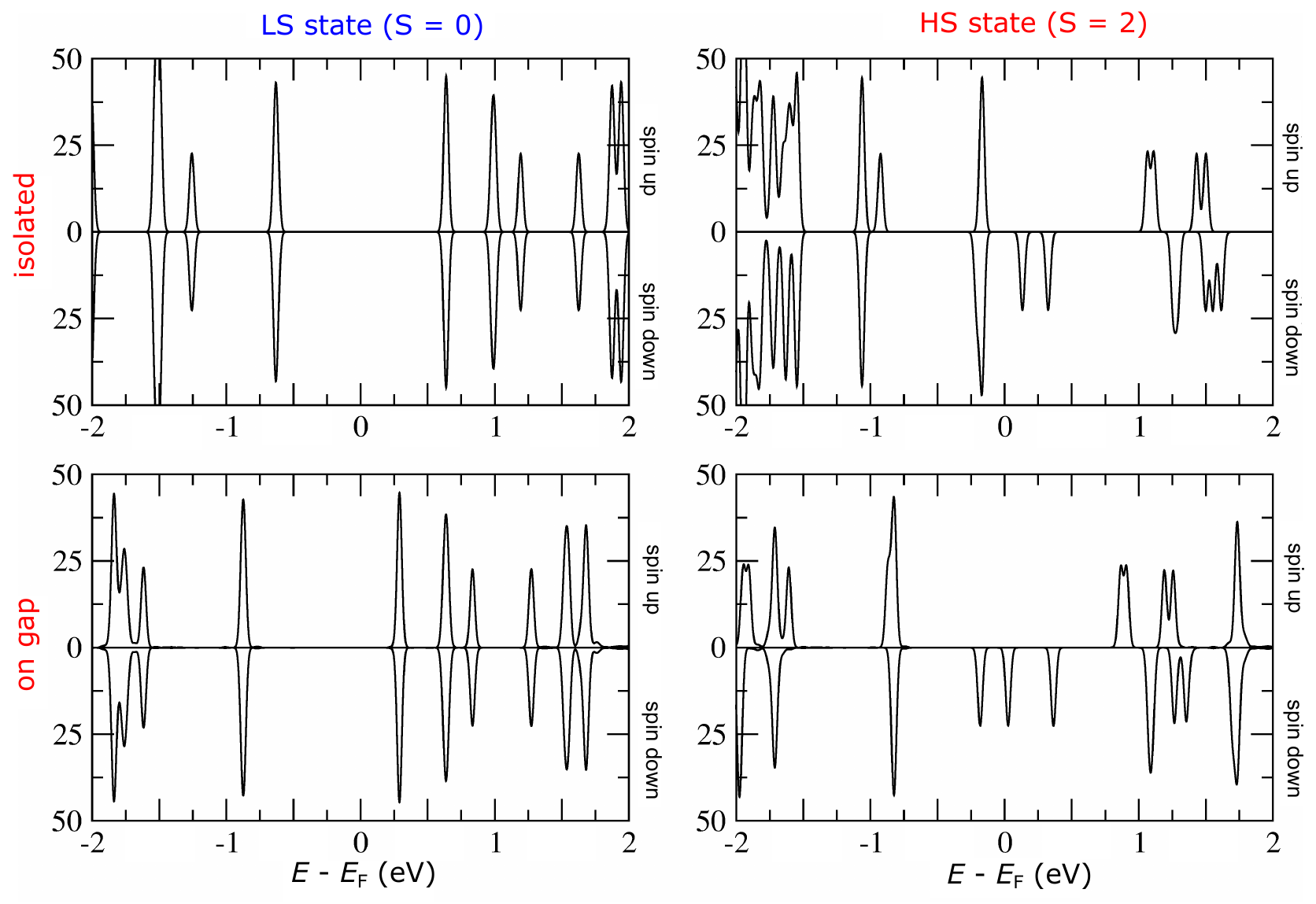}
		\caption{Comparison of the total projected density of states for the isolated molecule and the molecule deposited on the gap in the HS and LS states. The molecular orbitals associated to Fe undergo mainly a rigid displacement in energy.}
		\label{FigureS35}
	\end{figure}
	
	The local density of states of the orbitals closer to the Fermi level is plotted in Figure \ref{FigureS34}. The 2, 6, 7 and 8 states corresponding to d$_{z^2}$, d$_{yz}$, d$_{xz}$ and d$_{xy}$ respectively, are slightly delocalized around the Fe core. State 4, with a very small contribution of d$_{yz}$ is very delocalized in the central part of the molecule up to the oxygen atoms. The main contribution of d$_{x^{2}-y^{2}}$ lies at higher absolute energies above and below the Fermi level. The states 1, 3 and 5 are mainly present in the pyrene anchoring groups.

	The discussion and conclusions derived for the isolated molecule can be applied to the molecule placed linking the graphene electrodes. The main effect of the electrodes is a rigid shift of the molecular orbitals associated to the Fe to lower energies as shown in the comparative Figure \ref{FigureS35}.
	
	\subsection{Current switching mechanism and transition rates}
	
	We have identified two possible sources of the Low  to High Conductance switching, both due to Franck-Condon physics.
	The first (FCI) is the conventional switching mechanism for Fe(II) spin crossover molecules, where the ground state switches between S=0 and S=2 multiplets
	due to Temperature, light or perhaps the voltage \cite{Buhks1980}. The second corresponds to sequential ($N\rightleftharpoons N+1$) or 
	cotunneling ($N\rightleftharpoons N$) Franck-Condon physics (FCII) 
	\cite{Burzuri2014,Oppen2006}, corresponding to $S=0 \rightleftharpoons S=1/2$ or $S=2\rightleftharpoons S=3/2$, respectively. 
	
	\begin{figure}[h]
		\centering
		\includegraphics[width=0.50\columnwidth]{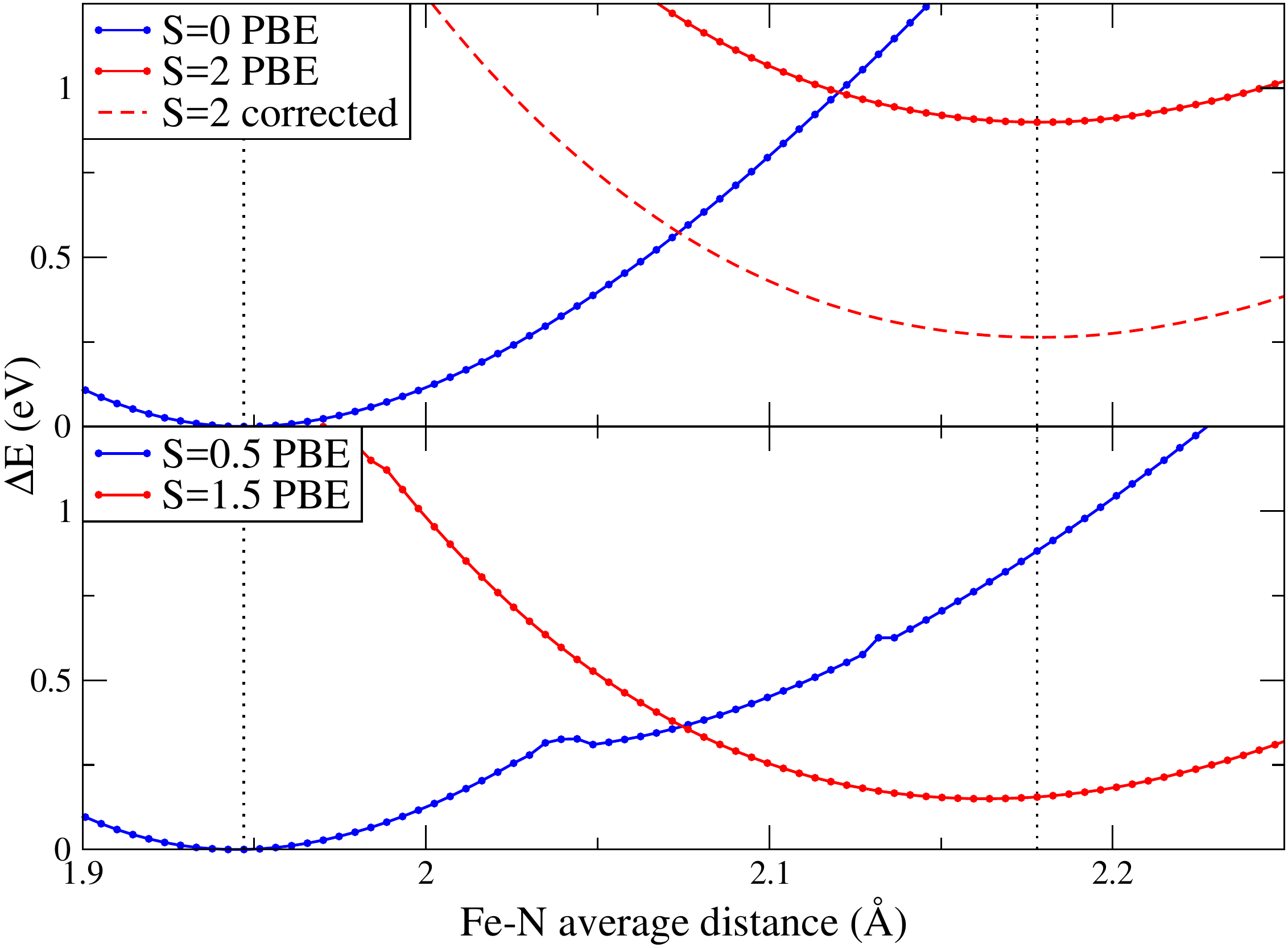}
		\caption{Energy difference between different spin states as a function of the Fe-N distance for the system with $N$ (top) and $N+1$ (bottom) electrons.}
		\label{fig:deltaE}
	\end{figure}
	
	To try and shed some light on this issue we plot in Figure \ref{fig:deltaE} the relative energy of the different spin solutions for the system with $N$ (top) and $N+1$ (bottom) 
	electrons, for structural solutions ranging between the ground state of the $S=0$ and the $S=2$ systems.
	The extra electron of the $N+1$ systems is located in a molecular orbital with contribution from the central Fe ion, thus the spin solutions correspond to
	the distribution of 7 $d$ electrons distributed in a strong ($S=0.5$) or weak ($S=1.5$) $O_h$ (or close to $O_h$) ligand field. We find that the Fe-N bond
	distance does not change by adding an extra electron. Therefore, the Franck-Condon factor for $N\rightleftharpoons N+1$ sequential tunneling is
	close to 1. This means that Franck-Condon blockade is not the dominant mechanism, so the FCII scenario is ruled out.
	On the other hand, there is a strong difference in the Fe-N bond distance of the $S=0$ and $S=2$ systems indicating a strong 
	spin-phonon coupling. This fact implies that
	Franck-Condon factors for spin-switching are small as usually happens with Fe(II) SCO molecules. We analyse therefore whether the FCI scenario fits with our
	experiments. 
	
	\begin{figure} 
		\centering
		\includegraphics[width=0.50\columnwidth]{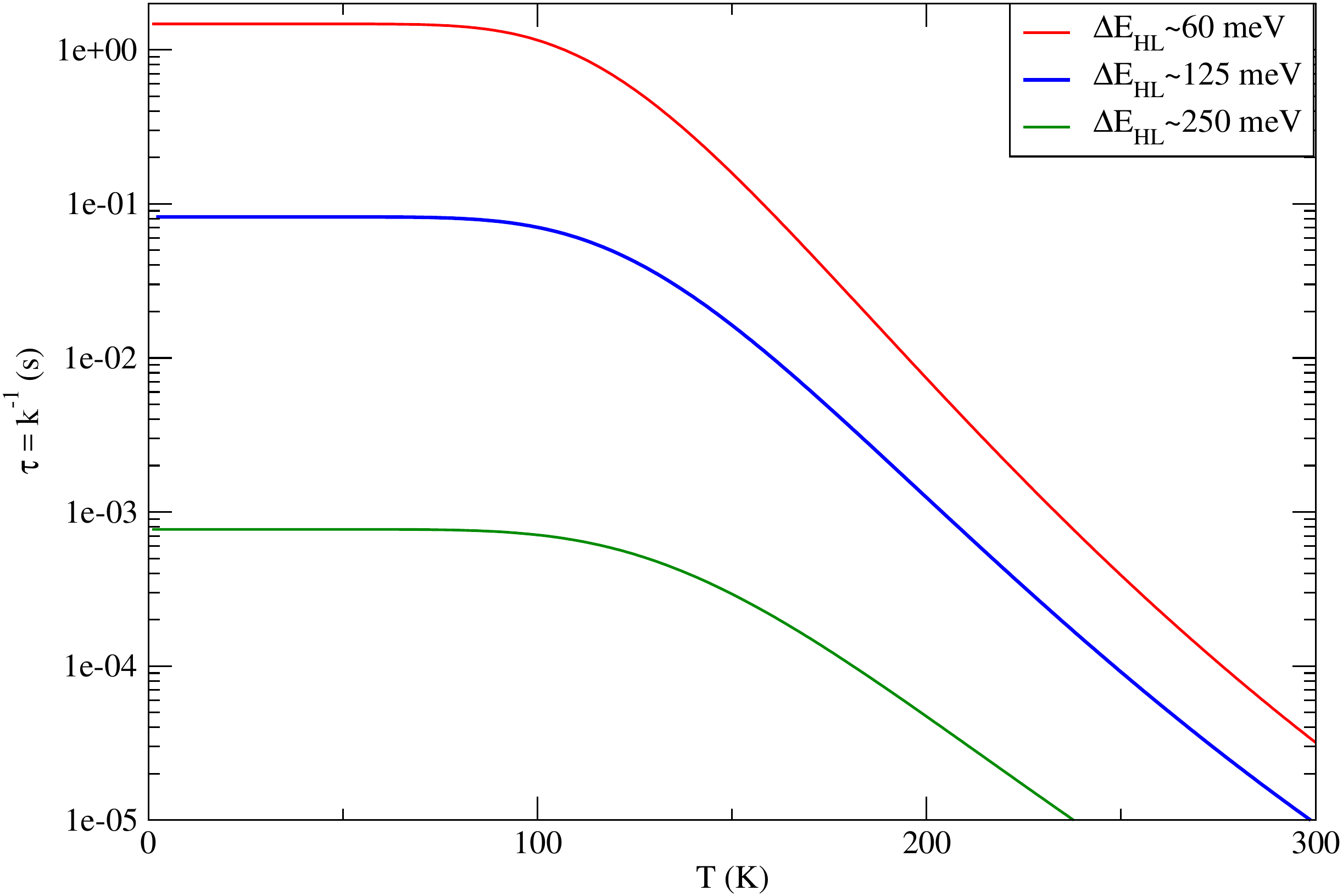}
		\caption{Average lifetime $\tau$ of the spin states as a function of the temperature for different values of $\Delta$E$_{HL}$ }
		\label{fig:rate}
	\end{figure}

	Literature tells that Fe-Terpyridine molecules tend to show slower switching times than other Fe(II) SCO molecules, and that 
	those switching times are temperature-independent up to higher threshold temperatures
	\cite{Buhks1980,thesis2014}.
	We borrow the transition rate discussion for $S=0\rightleftharpoons S=2$ switching  from Buhks {\it et al.}\cite{Buhks1980} :
	
	\begin{equation}
	k=\frac{2\pi }{\hbar}V_{SO}^2\,G
	\end{equation}
	
	\noindent
	where $V_{SO}$ indicates the second order transition between electronic states of $S=0$ and $S=2$ due to the spin-orbit 
	coupling and $G$ is the Franck-Condon factor which measures the vibrational overlap. To give rough numbers, 
	we use the energy difference between the $S=2$ and $S=0$ ground states $\Delta$E$_{LH}$, that
	we extract from our DFT GGA calculations; we find it to be 900 meV (7260 cm$^{-1}$).
	However, we note that DFT GGA calculations are known to overstabilize the LS state, thus overestimating $\Delta$E$_{LH}$. Indeed, our multielectronic calculations 
	of the core Fe$^{2+}$ levels give an  estimate $\Delta$E$_{LH}$ $\sim$250 meV (2000 cm$^{-1}$).  So we have decided to shift the two parabolae in Figure (\ref{fig:deltaE}) accordingly. We can also estimate from the curvature of the two parabolae an elastic constant $K=240$ N/m. The corresponding vibrational 
	energy is $\hbar\omega$=67 meV (537 cm$^{-1}$). From these parameters we calculate the transition rate as a function of the temperature in 
	Fig. (\ref{fig:rate}). For our estimated value of $\Delta$E$_{HL}$ $\sim$250 meV we find a constant transition rate up to more than 100 K, with a value 
	of $\tau$ $\sim$10$^{-3}$ s. This transition rate can be strongly modified if the deposition of the molecule modifies the energy
	difference between spin states to $\Delta$E$_{HL}$ $\sim$125 meV ($\tau$ $\sim$10$^{-1}$ s) or $\Delta$E$_{HL}$ $\sim$60 meV ($\tau$ $\sim$1 s), for example. 
	Other factors, such as the changes in the vibrational states due to the deposition of the molecule, can also play a role.

\clearpage
\bibliography{SCO_arxiv} 
\bibliographystyle{rsc} 

\end{document}